\documentclass[a4paper, preprint]{jpconf}
\usepackage{graphicx}
\usepackage{epsfig}
\usepackage[caption=false]{subfig}
\usepackage{amsmath}
\usepackage{amsfonts}
\usepackage{amssymb}
\usepackage{float}
\usepackage{color}
\usepackage{graphicx}
\usepackage{graphics}
\usepackage{hyperref}
\usepackage{adjustbox,array}
\hypersetup{}
\usepackage[utf8x]{inputenc} 
\usepackage{epstopdf}
\usepackage{multirow}
\usepackage{slashed}
\usepackage{booktabs}
\usepackage{gensymb}
\usepackage{comment}
\begin{document}
%\flushright{\Large KIAS-Q21012} \\
%%%%%%%%%%%%%%%%%%%%%%%%%%%%%%%%%%%%%%%%%%%%%%%%%%%%%%%%
\title{Studying QCD modeling of uncertainties in particle spectra from dark-matter annihilation into jets}
%%%%%%%%%%%%%%%%%%%%%%%%%%%%%%%%%%%%%%%%%%%%%%%%%%%%%%%%

\author{Adil Jueid}
\address{Quantum Universe Center, Korea Institute for Advanced Study, Seoul 02455, Republic of Korea}
\address{School of Physics, Konkuk University, Seoul 02095, Republic of Korea}
\ead{adiljueid@kias.re.kr}

\begin{abstract}
Motivated by various excesses observed by the Fermi-LAT and AMS collaborations, we perform a detailed analysis of QCD uncertainties on particle spectra from dark-matter annihilation (or decay) into jets. When annihilated to SM particles, the final-state annihilation products undergo various complicated processes such as QED and QCD bremsstrahlung, hadronisation, and hadron decays. These processes contain some intrinsic uncertainties which are usually difficult to model and which are neglected in physical analyses. First, we perform several re-tunings of the fragmentation function parameters. Then, we estimate two kinds of uncertainties: {\it (i)} perturbative from QCD showers and {\it (ii)} non-perturbative from hadronisation function. The results are tabulated for a wide range of dark matter masses, $m_\chi \in [10, 10^5]~{\rm GeV}$, and annihilation channels. They can be found on Zenodo: \url{https://doi.org/10.5281/zenodo.3764809}.
\end{abstract}

%\preprint{KIAS-Q21012} 

%%%%%%%%%%%%%%%%
\section{Introduction}
%%%%%%%%%%%%%%%%

Observations at the inter-galactic, galactic and cosmological scales provide strong evidence for the existence of Dark Matter (DM) in the universe. The measurements of the cosmological scale structure favour the so-called cold DM (CDM). In this scenario DM was not relativistic in the era of structure framework, and as the temperature drops its density decouples (this is known as the thermal freeze-out scenario). In particle physics framework, this scenario can be accommodated by extending the Standard Model (SM) with weakly interacting massive particles (WIMPs). This scenario has driven numerous efforts to test it through direct detection, indirect detection and collider experiments. Indirect detection experiments such as the Fermi Large Area Telescope (LAT) \cite{Fermi-LAT:2009ihh}, AMS \cite{AMS:2013fma} or IceCube \cite{Aartsen:2012kia} provide a promising avenue to detect WIMPs. Final state particles, such as  photons, positrons, neutrinos, or anti-protons, are produced from either the annihilation \cite{Griest:1990kh, Bergstrom:2000pn}, co-annihilations \cite{Baker:2015qna}, or decays \cite{Azri:2020bzl} of WIMPs. An excess of events above the SM backgrounds has been observed recently in the gamma-ray spectra by the Fermi-LAT collaboration \cite{TheFermi-LAT:2015kwa}. This is excess, called the Galactic Center Excess (GCE), seems to be originated from dark-matter annihilation; see e.g. \cite{Goodenough:2009gk}. Besides, several phenomenological studies have been carried out to address the GCE within particle physics models, {\it e.g.,} the MSSM \cite{Caron:2015wda, Bertone:2015tza, Butter:2016tjc, Achterberg:2017emt}. One of the most important outcome of these analyses is that the quality of the fits to the GCE depend on the precision of the theoretical predictions used to draw the gamma-ray spectra and that it is very important to have the theoretical uncertainties under control \cite{Caron:2015wda}.  The most dominant source of stable particles in DM annihilation/decay processes is Quantum Chromodynamics (QCD) jet fragmentation. Final state stable particles such as photons or positrons are then produced from a sequence of complex processes that include QED and QCD bremsstrahlung, hadronisation, and hadron decays. Unfortunately, there are no first-principle solutions to the problem of color fragmentation. We note that only phenomenological models such Fragmentation Functions  \cite{Metz:2016swz} or explicit dynamical models such as the string~\cite{Artru:1974hr,Andersson:1983ia} or cluster~\cite{Webber:1983if,Winter:2003tt} models are the up-to-date solutions to the hadronisation problem. The property of jet universality (or factorisation) tells us that the hadronisation process factorises from the short-distance hard-scattering processes. Therefore, one can determine the parameters of the hadronisation model from {\it e.g.} fit to LEP data (in $e^+ e^- \to {\rm hadrons}$) and use them to make predictions for DM annihilation. An interesting question concerns the size of the uncertainties on the theoretical predictions. It was found that different MC event generators have excellent agreement in both the peak as well as in the bulk of the spectra while they can have very large differences in the tails \cite{Cembranos:2013cfa}.  We stress out that the excellent level of agreement between the different MC event generators is simply due to the fact that they are tuned to the same set of data comprising of LEP measurements at the $Z$-boson pole \cite{Buckley:2009bj,Buckley:2010ar,Skands:2010ak,Platzer:2011bc,Karneyeu:2013aha,Skands:2014pea,Fischer:2014bja,Fischer:2016vfv,Kile:2017ryy,Reichelt:2017hts}. To assess this, we perform a comprehensive analysis of QCD uncertainties on both the fragmentation function parameters and the perturbative shower model using an alternative method and within the same model wherein the default Monash 2013 tune~\cite{Skands:2014pea} of the \textsc{Pythia} version 8.2.35 event generator~\cite{Sjostrand:2014zea} is used as our baseline. The new tunings are performed using a selection of experimental measurements from $e^+e^-$ colliders preserved in the \textsc{Rivet}~\cite{Buckley:2010ar} analysis package combined with the \textsc{Professor}~\cite{Buckley:2009bj} parameter optimisation tool. 

\begin{figure}[!t]
\centering
\begin{tabular}{c|cc}
QED bremsstrahlung & \multicolumn{2}{c}{QCD fragmentation and hadron decays}\\
\includegraphics[scale=0.65]{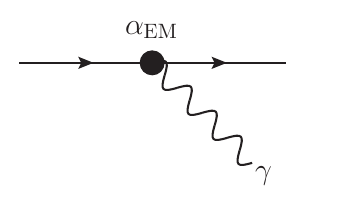} &
\includegraphics[scale=0.65]
{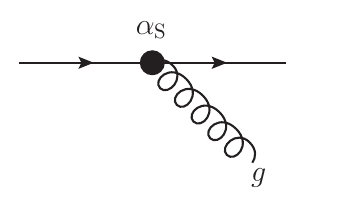} &
\raisebox{-2.5mm}{\includegraphics[scale=0.65]{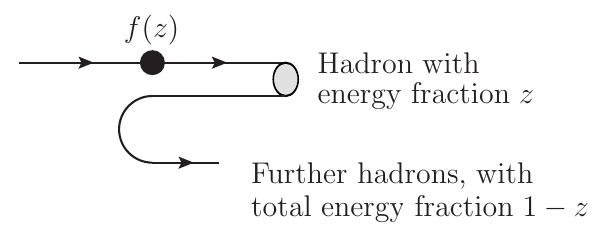}}\\
Dominates at high $x_\gamma$ & \multicolumn{2}{c}{Photons from $\pi^0\to\gamma\gamma$ dominate bulk (and peak) of spectra}
\end{tabular}
\caption{Illustration of the main parameters that affect the photon spectra ($x_\gamma = E_\gamma/m_\chi$) from DM annihilation into jets. Here we show the electromagnetic coupling $\alpha_\mathrm{EM}$ (left), the strong coupling $\alpha_S$ (middle), and the nonperturbative fragmentation function $f(z)$ (right). \label{fig:MCparams}}
\end{figure}

%%%%%%%%%%%%%%%%%%%%%%%%%%%
\section{Origin of Photons and Tunes}
%%%%%%%%%%%%%%%%%%%%%%%%%%%
In this section, we discuss briefly the physics modeling of photons in a generic DM annihilation, and the origin of photons (a more detailed discussion can be found in \cite{Amoroso:2018qga, Amoroso:2020mjm}). We consider a generic annihilation/decay process $\chi\chi \to \displaystyle\prod_{i=1}^{n} X_i$ followed by the decay $X_i \to \displaystyle\prod_{k=1}^{m} Y_{ik}$ (this factorisation relies on the narrow width approximation which is a good approximation here). If $X_i$ or its decay products $Y_{ik}$ contain photons or electrically charged particles, additional photons are produced via the branchings $X_i^\pm \to X_i^\pm \gamma$ with enhanced probabilities for both soft and collinear photons. Additional charged fermions are produced via the splittings $\gamma \to f\bar{f}$ with probabilities that are enhanced at very low virtualities. The rates of QED processes are determined by the value of the electromagnetic fine-structure constant, $\alpha_{\rm EM}$ (see Fig.~\ref{fig:MCparams}a). QCD bremsstrahlung occurs if the $X_i$ particles or $Y_{ik}$ are coloured particles. The modeling of the QCD showers is similar to the QED bremsstrahlung with the corresponding rates are governed by the effective value of the strong coupling constant $\alpha_S$ (see Fig.~\ref{fig:MCparams}b) which is evaluated at a renormalisation scale that is proportional to the shower evolution variable. We note that the value of $\alpha_S(M_Z^2)$ in \textsc{Pythia}~8 is different from the its value in the $\overline{\rm MS}$ scheme for two reasons {\it (i)} A set of universal corrections in the soft limit implies that the strong coupling should be defined in the CMW~\cite{Catani:1990rr} rather than the conventional $\overline{\mathrm{MS}}$ scheme and {\it (ii)} A good agreement between \textsc{Pythia8} and experimental measurements of $e^+ e^- \to 3~\mathrm{jets}$ \cite{Skands:2010ak,Skands:2014pea} is achieved through an increase of the value of $\alpha_S(M_Z)$ by about $10\%$. Finally, any produced coloured particle must be confined inside colourless hadrons through the hadronisation process which takes place at distance scale of order of the proton size $\sim 10^{-15}~{\rm m}$. The majority of photons are produced from the decays of neutral pions where the number and energy of these photons are strongly correlated with the predicted pion spectra. The description of this process is embedded in the \emph{fragmentation function}, $f(z)$, which gives the probability for a hadron to take a fraction $z \in [0,1]$ of the remaining energy at each step of the (iterative) string fragmentation process (see Fig.~\ref{fig:MCparams}c). We close this subsection noting that most of the photons are coming from pion decays followed by the contribution of $\eta$ decays which accounts for about $4\%$. The contribution of bremsstrahlung photons is very subleading and dominates in the  about $88$-$95\%$ depending on the annihilation channel and on the DM mass. 

In this study, we use \textsc{Pythia8} version 8.235 with the \textsc{Monash} tune~\cite{Skands:2014pea} used as baseline for the new parameter retunings. The parameters under investigations are those of the flavour-independent light quark fragmentation function: the $a$, $b$~(or $\langle z_\rho \rangle$) and $\sigma$ denoted by $\texttt{StringZ:aLund}$, $\texttt{StringZ:bLund}$~($\texttt{StringZ:avgZLund}$) and $\texttt{StringPT:sigma}$. The optimisation procedure is performed with the help of the \textsc{Professor} toolkit version 2.2 \cite{Buckley:2009bj} based on the implemented analyses in \textsc{Rivet}~2.5.4 \cite{Buckley:2010ar}. The \textsc{Professor} toolkit relies on a method that allows for a simultaneous optimisation of several parameters by using analytical expressions which approximate the true dependence of the MC response on the model parameters (the analytical expressions are cast as polynomials). To reduce the residuals -- differences between real MC response and the approximate formulas -- we employ a fourth-order polynomial. A standard $\chi^2$ minimisation procedure is used to obtain the values of the parameters at the minimum (please see \cite{Amoroso:2018qga} for more details). The results of the tunings are 
 \begin{eqnarray}
  \verb|StringZ:aLund|  &=& 0.5999\pm0.2, \nonumber \\
  \verb|StringZ:avgZLund| &=& 0.5278^{+0.027}_{-0.023}, \\
  \verb|StringPT:sigma| &=& 0.3174^{+0.042}_{-0.037} \nonumber
  \label{eq:tunes:results}
 \end{eqnarray}

%%%%%%%%%%%%%%%%%%%
\section{QCD uncertainties}
\label{tune:uncertainties}
\begin{figure}[!t]
\centering
\includegraphics[width=0.495\linewidth]{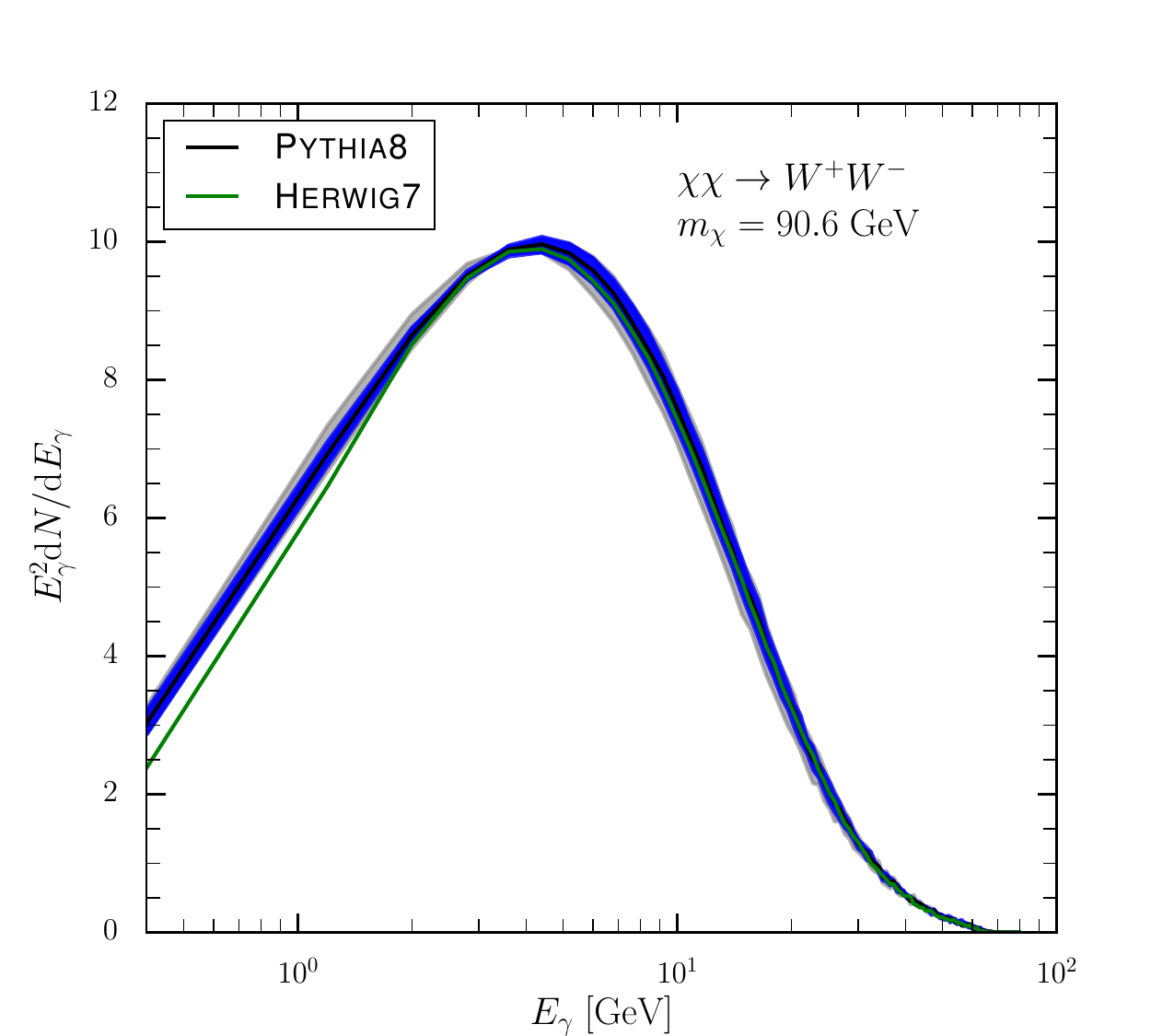}
\hfill
\includegraphics[width=0.495\linewidth]{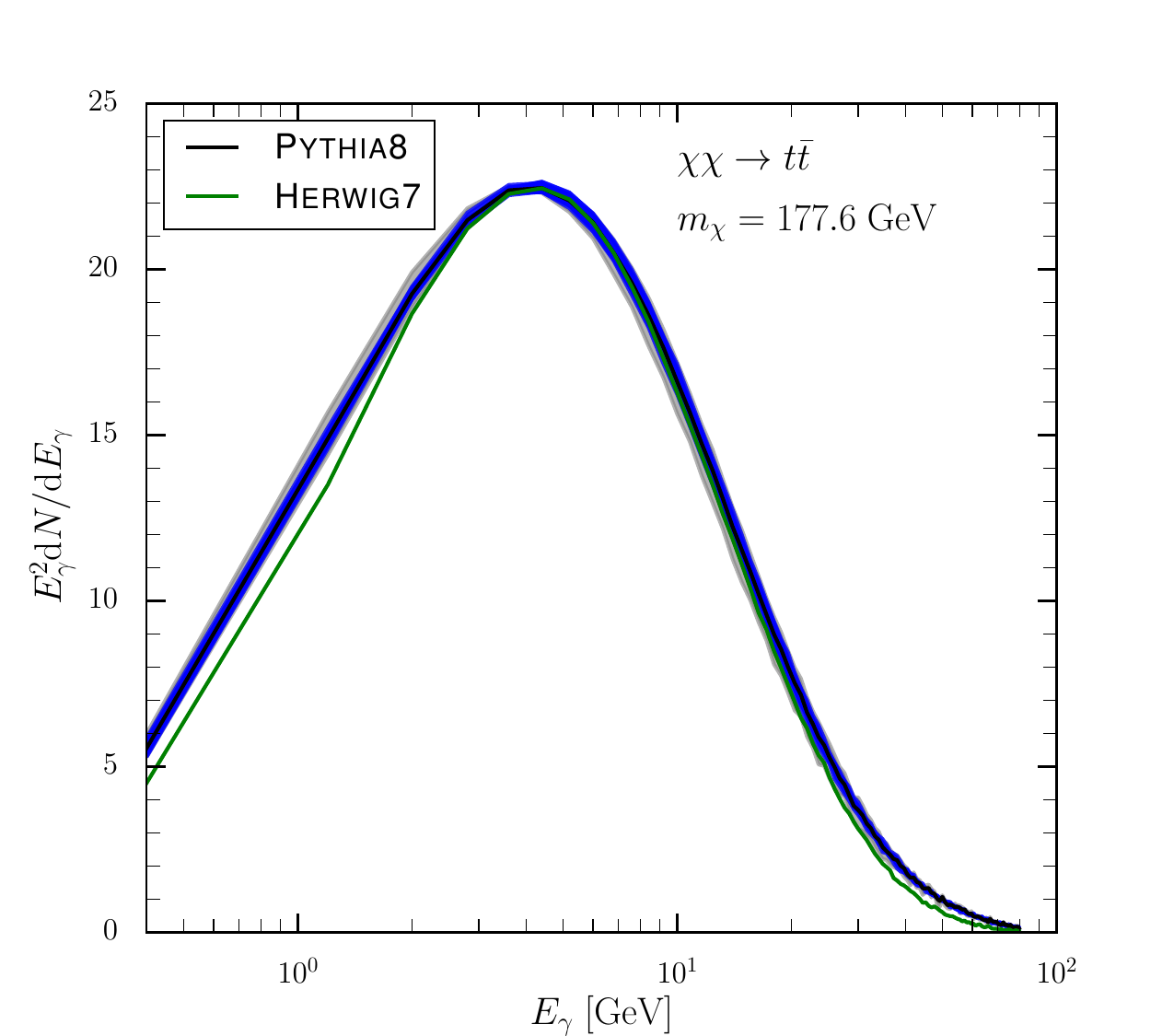}
\caption{The distribution of the energy spectra of gamma-rays at the production from dark matter annihilation into $W^+ W^-$ ({\it left}) and $t\bar{t}$ ({\it right}). Both the uncertainties from parton showering (gray bands) and from hadronisation (blue bands) are shown. Predictions from \textsc{Herwig7} version 7.1.3 are shown as a green solid line.}
\label{fig:DMimpact}
\end{figure}

%%%%%%%%%%%%%%%%%%%
The QCD uncertainties on the gamma-ray spectra are divided into two categories: {\it (i)} perturbative uncertainties on the shower evolution variable and {\it (ii)} non-perturbative uncertainties connected to the variations, around the nominal value, of the parameters of the hadronisation model. The uncertainties from the parton showers are estimated using an automatic method developed by the authors of \cite{Mrenna:2016sih}. The non-perturbative are determined from a weighted $\chi^2$ minimisation where we tune individually to fifteen measurements to get different best-fit points and combine those to get the $68\%~{\rm CL}$ errors on the parameters. To get a fully comprehensive envelope, we consider all the possible $N_{\rm param}^3 - 1$ variations (with $N_{\rm param} = 3$ are the parameters of the hadronisation model). The impact of the QCD uncertainties on the photon energy spectra from dark matter annihilation into $W^+ W^-$  with $m_\chi = 90.6$ GeV and $t\bar{t}$ with $m_\chi = 177.6$ GeV is shown in figure \ref{fig:DMimpact}.  These two values of the dark-matter masses are motivated from a study of the GCE using PASS8 data in the framework of the phenomenological-MSSM \cite{Achterberg:2017emt}. We can see that the hadronisation uncertainties are dominant in both the peak and the high-energy tail of the spectrum while the perturbative uncertainties are dominant in the rest of the spectrum (they can change the peak of the energy distribution). 

%%%%%%%%%%%%%%%%%%%%%%%%%%%%%%%%%%
\section{Public Data on Zenodo}
\label{sec:code}
%%%%%%%%%%%%%%%%%%%%%%%%%%%%%%%%%%
The main result of this study (documented in details in \cite{Amoroso:2018qga}) is provided in the form of datasets which can be found on \texttt{Zenodo} \cite{Amoroso:2019Zenodo}. The datasets have been produced for five stable final states: gamma rays, positrons, electron anti-neutrinos, muon anti-neutrinos and tau anti-neutrinos. The calculations of the flux at the production were done for eleven annihilation channels $\chi \chi \to e^+ e^-, \mu^+ \mu^-, \tau^+ \tau^-, q\bar{q}~(q=u,d,s), c\bar{c}, b\bar{b}, t\bar{t}, W^+ W^-, ZZ, gg, ~\mathrm{and}~hh$ (the results can be used for studies targeting DM decay instead of DM annihilation but care has to be taken regarding the phase space factors). The dark matter masses covers a very wide range from $5$ GeV to $100$ TeV. For each final state and annihilation channel, twelve tables are provide in \texttt{zip} format which correspond to the nominal value of the hadronisation model (denoted by 'AtProduction-Hadronization1-\$TYPE.dat'), the nine variations around the nominal value (denoted by 'AtProduction-Hadronization\$h-\$TYPE.dat' with $h=2,\cdots,10$) and the two variations of the renormalisation scale by a factor of $2$ in positive and negative directions (denoted by 'AtProduction-Shower-Var\$s-\$TYPE.dat', $s=1,2$). Our predictions have been compared to the results of the PPPC4DMID \cite{Cirelli:2010xx} and we found that there are some differences in the peak and the tail\footnote{The differences have been discussed in great details in appendix D of \cite{Amoroso:2018qga} while the correct method to generate the spectra is found in appendix C of the same paper.}.

%%%%%%%%%%%%%%%%%%%
\section{Conclusions}
%%%%%%%%%%%%%%%%%%%
In this talk, we discussed the results of a {\it first study} of the QCD uncertainties on particle spectra from DM annihilation. Using the most recent \textsc{Monash} tune as our baseline, we have performed several retunings of the parameters of the string hadronisation model. We then defined a set of parametric variations of these parameters and studied their impact on the gamma-ray spectra for two benchmark scenarios. The full datasets which can be considered as an update to the PPPC4DMID are now public on \texttt{Zenodo} and can be found in following link \url{http://doi.org/10.5281/zenodo.3764809}.

\section*{Acknowledgements}
The author would like to thank Simone Amoroso, Sascha Caron, Roberto Ruiz de Austri and Peter Skands for the useful discussions throughout the study that leads to the results presented in this talk. This work is supported in part by a KIAS Individual Grant No. QP084401 via the Quantum Universe Center at Korea Institute for Advanced Study and by the National Research Foundation of Korea, Grant No. NRF-2019R1A2C1009419. 

\section*{References}
\bibliography{bibliography.bib}

\end{document}